\documentclass[aps,final,notitlepage,oneside,superscriptaddress,noshowpacs,centertags]{revtex4-1}
\usepackage[english]{babel}
\usepackage{graphicx}
\usepackage{latexsym}
\usepackage{amssymb}
\usepackage{amsmath}
\usepackage{float}

\begin{document}
\title{Conformal invariance and phenomenology of cosmological particle production}
\author{V. A. Berezin}\thanks{e-mail: berezin@inr.ac.ru}
\affiliation{Institute for Nuclear Research of the Russian Academy of Sciences, 
	pr. 60-letiya Oktyabrya 7a, Moscow, 117312 Russia}
\author{V. I. Dokuchaev}\thanks{e-mail: dokuchaev@inr.ac.ru}
\affiliation{Institute for Nuclear Research of the Russian Academy of Sciences, 
	pr. 60-letiya Oktyabrya 7a, Moscow, 117312 Russia}
\affiliation{National Research Nuclear University ``MEPhI'', 
	Kashirskoe sh. 31, Moscow, 115409 Russia}
\author{Yu. N. Eroshenko}\thanks{e-mail: eroshenko@inr.ac.ru}
\affiliation{Institute for Nuclear Research of the Russian Academy of Sciences, 
	pr. 60-letiya Oktyabrya 7a, Moscow, 117312 Russia}

\date{\today}

\begin{abstract}
Starting with the idea to describe phenomenologically the particle creation in the strong gravitational fields, we introduced explicitly the particle number nonconservation (= creation law) into the action integral with the corresponding Lagrange multiplier. Following the fundamental result by Ya. B. Zel'dovich and A. A. Starobinsky (1977) we then postulated that the rate of particle creation is proportional to the square of Weyl tensor. Concerning the conformal invariance, yet another question arises: how the scalar field could know about the surgery made on the metric tensor (if such an invariance is the fundamental law of Nature and not just the mathematical exercise)? The only way is that the scalar field is itself the part of metric, namely, the conformal factor. We showed, that  such an identification results in the natural appearance of the quartic self-interaction term in the scalar field Lagrangian, which is needed to make particle massive. And it is just quartic, because our space-time is four-dimensional. 
\end{abstract}

\maketitle

\section{Instead of ``Introduction''} 

\begin{itemize}
\item 2015 --- 100 years Of the Einstein's  General Relativity \cite{AE,AE2}
\item 2017 --- 100 years of the Cosmology  \cite{AEcosm}.
\item 2018 --- 100 year of the Weyl geometry and Weyl Conformal Gravity \cite{Weyl}.
\end{itemize}

\section{Phenomenology of particle creation}

The quantum field theory predicts the phenomenon of particle creation from the vacuum fluctuations in the presence of the external fields. The modern cosmology based on the Friedmann`s model of the expanding Universe \cite{Friedmann,Friedmann24} provides us with the strong and rapidly varying gravitational field in the vicinity of the Big Bang singularity. In 1970's there was a big activity in studying the particle creation by the quantized scalar field on the background scalar field on the background metric of the homogeneous and slightly anisotropic cosmological models \cite{Parker69,Grib70,Zeld70,Zeld71,Zeld72,Parker73,Fulling73,Parker73b,Parker74,Parker74b,Parker74c,Lukash74}. Here we are interested in two main results of these investigations. First, it is the necessity of inclusion the additional terms into the gravitational  part of the total action integral. Namely, the terms which  are quadratic in Riemann curvature tensor and its counterparts, Ricci tensor and curvature scalar. And second, the role of particle production, as was shown by Ya.~B.~Zel'dovich  and A.~A.~Starobinsky \cite{Zeld77}, is proportional to the square of Weyl tensor, the latter being the completely traceless part of Riemann tensor. The appearance of the quadratic terms was foreseen by A.~D.~Sakharov \cite{Sakh} in 1967 (50 years ago!). However the attempts to solve the self-consistent problem, with the account for back reaction of the (averaged) energy-momentum tensor of the quantized scalar fields on the space-time metric encounter serious obstacles. The matter is that in order to solve the quantum part of the problem, one needs to impose the appropriate boundary conditions, but this would be done only after solving the gravitational equations, for which the averaged quantum fields serve as the source. That is why we have to use some phenomenological approach. Our goal is not only to take into account the energy-momentum tensor of the already created particles, but also the reaction of space-time structure on the very process of the matter creation. There is a hope that this may be helpful in solving the singularity problem by violating the energy momentum-condition. 
  
We are using the hydrodynamical description of the particle content elaborated by J. R. Ray \cite{Ray}. The details can be found also in \cite{Ber87} and \cite{bde16}. The convenience of chosen approach is that the particle number conservation law infers the hydrodynamical action integral explicitly as the constraint with corresponding Lagrange multiplier $\lambda_1$. And we simply replace it with the particle number creation law, namely
\begin{equation}
\int \lambda_1(nu^\sigma)_{;\sigma}\sqrt{-g}\,dx \quad  \rightarrow \quad 
\int \lambda_1[(nu^\sigma)_{;\sigma}-\beta C^2]\sqrt{-g}\,dx,
\label{beta}
\end{equation}
where we already inserted the square of the Weyl tensor, $C^2$, exploring the already mentioned fundamental result \cite{Zeld77}  for the cosmological particle creation. It is quite interesting  to note that the Lagrange multiplier $\lambda_1$ is defined, actually, up to the additive constant. Indeed, let us replace $\lambda_1\rightarrow\lambda_1+\gamma_0$, $\gamma_0=const$. Then, $\gamma_0 (nu^\sigma)_{;\sigma}\sqrt{-g}=\gamma_0 (nu^\sigma)_{,\sigma}$ is the full derivative which does not change the equation  of motion, while the term $-\gamma_0\beta C^2=\alpha_0 C^2$ is just the Lagrangian density of the Weyl conformal gravity. And this is in spirit of the Sakharov's idea that the gravitational field is not fundamental, but simply the tensions of quantum vacuum fluctuations of all the matter fields.

And what about the scalar field, $\chi$, the source of particle creation? We will choose for it the simplest possible Lagrangian which gives the linear equations of motion (without any self-interactions),
\begin{equation}
{\cal L}_{\rm scalar}=\frac{1}{2}\chi_\sigma\chi^\sigma-\frac{1}{2}m^2\chi^2,
\quad \chi_\sigma=\chi_{,\sigma}, \quad 
\chi^\sigma=g^{\sigma\lambda}\chi_{\lambda}.
\label{scalar}
\end{equation}
Note that it is not that scalar field which is to be quantized, but some its residual part, because the already created particles are just the quanta of the genuine scalar field.

\section{Conformal invariance}

The conformal gravity was invented by H. Weyl in 1918 \cite{Weyl}. Then it was recognized that it allows only massless particle to exist, and on this ground the theory was rejected. But nowadays, such an unpleasant feature can be ``corrected''  by the Brout-Englert-Higgs mechanism for the spontaneous symmetry breaking \cite{tHooft}. Besides, the vacuum space-time with very high symmetry is a good candidate for the creation of the universe from ``nothing'' \cite{Vilenkin}. The idea that the initial state of the universe should be conformally invariant is advocated also by R.~Penrose \cite{Penrose1,Penrose2}. We will consider the conformal invariance as the fundamental postulate.

By the conformal transformation we will understand the space-time depending scaling of the metric tensor $g_{\mu\nu}$:
\begin{equation}
ds^2=g_{\mu\nu}(x)dx^\mu dx^\nu=\Omega^2(x)\hat g_{\mu\nu}(x)dx^\mu dx^\nu
=\Omega^2(x)d\hat s^2.
\end{equation}
The, local, conformal invariance means $\delta S_{\rm tot}/\delta \Omega=0$. Therefore, we can consider the conformal factor $\Omega$ as an additional (to $g_{\mu\nu}$) dynamical variable. 

It is known for a long time that, if the conformal transformation of the metric tensor (written above) is supplemented by the transformation $\hat\chi=\Omega\chi$ for the massless scalar field, then, by adding the term $\chi^2R/12$, to the Lagrangian (\ref{scalar}) with $m^2=0$, one makes the latter conformal invariant (of course, up to the full derivative which is usually neglected). In this way the Einstein-Hilbert Lagrangian (modified by the ``dilaton'' field) appeared in the total action integral. Remember, that we started with the pure matter (hydrodynamical) term and demanded the particle creation law, from which the Weyl tensor square was extracted (again the Sakharov's idea). But, even for massless scalar field (i.\,e., when $m^2=0$) one serious problem is remained and should be somehow solved. It is the problem of the common sign in front of the curvature scalar $R$ and the kinetic term $\chi^\sigma\chi_\sigma$. Indeed, if one chooses the ``correct'' sign, i.\,e., $-\chi^2R/12$, in order to get the attractive Einstein gravity, then the kinetic term for the scalar field appeared to be wrong, and vice versa. Our choice is the ``correct'' sign for the gravity. This requires some explanation. First, we do not care about the ``correct'' sign for the kinetic term, because our scalar filed $\chi$ is not the genuine (i.\,e., fundamental) one: some part of it we have already ``used'' as the created particles. Second, the ``wrong'' sign is, in a sense, good since it allows even number of particles to be created.

One more thing. We are now working  in the framework of the Riemann geometry, This means that the only fundamental geometric quantity is the metric tensor $g_{\mu\nu}$, the connections $\Gamma_{\mu\nu}^\lambda$ being constructed exclusively from $g_{\mu\nu}$ and its derivatives (as well as  the Riemann curvature tensor $R_{\nu\lambda\sigma}^\mu$ and its contractions, $R_{\mu\sigma}$ and $R$). All the matter fields are considered as the gravitating sources and something external. Then, how the field $\chi$ could ``know'' that it must be conformally transformed following the transformation of metric tensor? Only, if it is the part of it. For the specific conformal factor $\Omega=\phi$ one can make $\hat\chi=\phi/l$, where $l$ is some factor with the dimension of length. Then, the scalar part of the total action takes the form (with the massive term restored)
\begin{equation}
S_{\rm scalar}=-\frac{1}{l^2}\int\left(\frac{1}{2}\phi^\mu\phi_\mu+\frac{R}{12}\phi^2
-\frac{\Lambda}{6}\phi^4\right)\sqrt{-\hat g}\,dx,
\end{equation}
where we introduced the cosmological term $\Lambda=3m^2$. We see that there appears the self-interaction term $\phi^4$, badly needed in order to switch on the Brout-Englert-Higgs mechanism! And the power $4$ in this term is only in the case of the four-dimensional space-time!

\section{In place of ``Conclusion'' and ``Discussion''}

First of all, we would like to pay tribute to P. I. Fomin whose paper ``Gravitational instability of vacuum and the cosmological problem'', published in 1973 \cite{Fomin}, made him the precursor not only of the idea of the quantum birth of the Universe (from ``nothing''), but also of the ``emergent'' universe scenarios \cite{Brout1,Brout2,Prig1,Prig2}. One of us (V.B) expresses gratitude to G. Bisnovatyi-Kogan for reminding about this \cite{BisnK}.

In 1918 Hermann Weyl introduced the the great physical idea \cite{Weyl}. If there exists some local transformation that leaves the action integral (and equations of motion)  invariant, then there should exist some field that propagates the knowledge about such a transformation from one point to another, He developed this idea for electrodynamics and much later R.~Utiyama extended it on the more general, nonabelian gauge field \cite{Utiyama}. Practically, this idea was realized in replacing the metric connections by new ones which included in addition to the metric tensor $g_{\mu\nu}$ and its derivatives, some vector (gauge) field $A_\mu$ combined with the metric tensor and Kronecker symbols. So, the geometry is no longer Riemannian, it is called the Weyl geometry. Of course, the curvature tensor being dependent solely on the connection, is changed (as well as the Ricci tensor and curvature scalar). And the Totla Lagrangian gets the new term, $F_{\mu\nu}F^{\mu\nu}$, where $F_{\mu\nu}$ is the gauge invariant strength of the vector field $A_\mu$. 

The transition from the Riemann to the Weyl geometry open new possibilities for us. Now we are not forced to identify the (residual vacuum) scalar field $\chi$ with the conformal factor. Instead, we can introduce (somehow) the interaction of this scalar field with the new gauge field (say, by adding the term $g^{\mu\nu}A_\mu\chi_{,\nu}$ to the Lagrangian). But, the our scalar field is not the only matter field, it is not exceptional at all. It looks much more logical and self-consistent to arrange the interaction between vector field $A_\mu$ and the energy flow density $E^\mu$, constructed from the total energy-momentum tensor $T_{\mu\nu}$. The problem is that $E^\mu=T^{\mu\nu}\xi_\nu$ depends crucially on the world line $\xi^\mu$ of the observer. Thus following such an idea, we would arrive at the observer-dependent theory (like quantum mechanics). Moreover, in addition to the equation of motion one should impose the consistency conditions, i.\,e., that the energy-momentum tensor entering the above mentioned term is the same as that obtained by varying the matter part of the total action integral with respect to the metric tensor.

\section*{Acknowledgments}

Authors acknowledge G. Bisnovatyi-Kogan,  A. Smirnov and A. Starodubtsev for helpful discussions. This research supported in part by the RFBR under grant No. 15-02-05038-a.

\end{document}